\documentclass[conference]{IEEEtran}
\IEEEoverridecommandlockouts

\usepackage{cite}
\usepackage{amsmath,amssymb,amsfonts}
\usepackage{algorithmic}
\usepackage{graphicx}
\usepackage{textcomp}
\usepackage{xcolor}
\usepackage{booktabs}
\usepackage{hyperref}
\usepackage{subcaption}
\usepackage{multirow}

\begin{document}

\title{DeepFake Forensics AI: A Multi-Modal Detection and 
Blockchain-Anchored Evidence Management Platform}

\author{
\IEEEauthorblockN{Naisha Minnah}
\IEEEauthorblockA{
Department of Computer Science\\
Providence Women's College, Calicut\\
Affiliated to University of Calicut, Kerala, India\\
naishaminnah@gmail.com}
}

\maketitle

\begin{abstract}
The proliferation of AI-generated synthetic media poses a critical 
threat to the integrity of digital evidence in legal and forensic 
contexts. Existing deepfake detection systems typically address a 
single modality and provide no mechanism for tamper-proof evidence 
preservation. We present \textbf{DeepFake Forensics AI}, a unified 
platform that detects synthetic media across image, video, and audio 
modalities, identifies generative architecture fingerprints, and 
anchors forensic evidence immutably on the Ethereum blockchain. 
Our system trains four independent neural networks from scratch: 
an EfficientNet-B4 image detector (AUC = 0.9868), a Bidirectional 
LSTM video detector (AUC = 0.9628), an ECAPA-TDNN audio detector 
(EER = 18.63\%), and a novel GAN fingerprinting module (accuracy = 
99.88\%) that identifies the generative architecture behind a fake 
image. Evidence files are hashed with SHA-256, stored on IPFS via 
Pinata, and registered on-chain via a Solidity smart contract with 
role-based access control. The platform provides a React frontend 
and FastAPI backend suitable for deployment in forensic and legal 
workflows. To our knowledge, this is the first system to unify 
multi-modal deepfake detection with blockchain-based chain-of-custody 
management.
\end{abstract}

\begin{IEEEkeywords}
deepfake detection, GAN fingerprinting, blockchain, forensic evidence, 
ECAPA-TDNN, EfficientNet, multimodal, IPFS, smart contracts
\end{IEEEkeywords}

\section{Introduction}
\label{sec:intro}

The rapid advancement of generative adversarial networks (GANs) and 
diffusion models has made the synthesis of photorealistic fake media 
trivially accessible. Deepfake videos, cloned voices, and GAN-generated 
images now circulate at scale across social media, news platforms, and 
increasingly, legal proceedings. Courts in multiple jurisdictions have 
already encountered challenges authenticating digital evidence suspected 
of AI manipulation~\cite{tolosana2020deepfakes}, exposing a critical gap 
between the pace of synthetic media generation and the forensic tools 
available to counter it.

Existing deepfake detection systems suffer from two fundamental 
limitations. First, they are predominantly unimodal --- addressing 
either visual or audio forgeries in isolation --- leaving adversaries 
free to exploit the unguarded modality. Second, they provide no 
mechanism for preserving detected evidence in a tamper-proof, 
legally admissible form. A forensic analyst who identifies a deepfake 
today has no cryptographically verifiable record of that finding.

This paper addresses both limitations with \textbf{DeepFake Forensics 
AI}, a unified platform built from the following contributions:

\begin{itemize}
    \item A \textbf{multi-modal detection pipeline} combining 
    frame-level image detection (EfficientNet-B4), temporal video 
    analysis (Bidirectional LSTM over EfficientNet-B4 embeddings), 
    and audio spoof detection (ECAPA-TDNN), all trained from scratch 
    on publicly available benchmarks.
    
    \item A \textbf{GAN fingerprinting module} that identifies the 
    specific generative architecture (ADM, BigGAN, Glide, or VQDM) 
    behind a fake image, achieving 99.88\% accuracy on a 4,800-image 
    held-out test set.
    
    \item A \textbf{GAN inversion module} using BigGAN and StyleGAN2 
    latent projection to reconstruct the latent vector of a fake image, 
    providing evidence of generative origin for court use.
    
    \item A \textbf{blockchain evidence registry} implemented as a 
    Solidity smart contract on Ethereum, combining SHA-256 content 
    hashing with IPFS decentralised storage and role-based access 
    control for forensic analysts and legal authorities.
\end{itemize}

To our knowledge, DeepFake Forensics AI is the first system to 
integrate multi-modal deepfake detection with cryptographically 
immutable chain-of-custody management, addressing the full forensic 
workflow from detection to court-admissible evidence preservation.

The remainder of this paper is organised as follows. Section~\ref{sec:related} 
reviews related work. Section~\ref{sec:architecture} describes the 
system architecture. Section~\ref{sec:method} details the methodology 
for each component. Section~\ref{sec:results} presents experimental 
results. Section~\ref{sec:conclusion} concludes.

\section{Related Work}
\label{sec:related}

\subsection{Deepfake Detection}

Early deepfake detection relied on hand-crafted features such as 
eye blinking patterns and facial landmark inconsistencies. The 
introduction of FaceForensics++~\cite{rossler2019faceforensics}, 
a large-scale benchmark covering five manipulation types (Deepfakes, 
Face2Face, FaceSwap, FaceShifter, and NeuralTextures), enabled 
systematic evaluation of learned detectors. Subsequent work 
demonstrated that CNN-based classifiers trained on compressed video 
frames generalise well across manipulation types~\cite{wang2019cnngenerated}. 
EfficientNet architectures~\cite{tan2019efficientnet} have since 
become a standard backbone for frame-level detection owing to their 
favourable accuracy-efficiency trade-off.

Video-level detection requires temporal reasoning beyond what 
frame-level classifiers provide. Recurrent architectures, particularly 
bidirectional LSTMs, have been applied to aggregate frame-level 
features across time, capturing inconsistencies in facial dynamics 
that single-frame models miss. Celeb-DF~\cite{li2020celeb} introduced 
a more challenging benchmark with higher-quality synthesis, exposing 
the generalisation limitations of models trained exclusively on 
FaceForensics++.

Audio deepfake detection has been benchmarked primarily on the 
ASVspoof challenge series~\cite{liu2019asvspoof}. ECAPA-TDNN~\cite{desplanques2020ecapa}, 
originally proposed for speaker verification, has demonstrated 
strong performance on anti-spoofing tasks owing to its emphasis on 
channel attention and multi-scale feature aggregation.

\subsection{GAN Fingerprinting}

The observation that CNN-generated images contain subtle spectral 
artefacts attributable to the specific architecture used for 
generation motivates GAN fingerprinting~\cite{wang2019cnngenerated}. 
Unlike binary real/fake detection, GAN fingerprinting frames the 
problem as multi-class classification over generative architectures. 
High-pass filtering as a preprocessing step has been shown to 
amplify these artefacts, improving classification accuracy. Our 
work extends this direction with a two-stage pipeline: binary 
real/fake screening followed by four-class GAN type identification 
(ADM, BigGAN, Glide, VQDM).

\subsection{Blockchain for Digital Forensics}

Distributed ledger technologies offer a natural solution to the 
chain-of-custody problem in digital forensics. Immutable transaction 
records on public blockchains~\cite{nakamoto2008bitcoin} provide 
cryptographic proof of existence and integrity for digital artefacts. 
Prior work has explored blockchain-based evidence management for 
general digital forensics, but to our knowledge no existing system 
integrates deepfake detection with on-chain evidence registration. 
Our platform addresses this gap by combining SHA-256 content hashing, 
IPFS decentralised storage, and an Ethereum smart contract with 
role-based access control, creating an end-to-end forensic workflow 
from detection to court-admissible evidence preservation.

\section{System Architecture}
\label{sec:architecture}

\begin{figure}[htbp]
\centering
\includegraphics[width=\columnwidth]{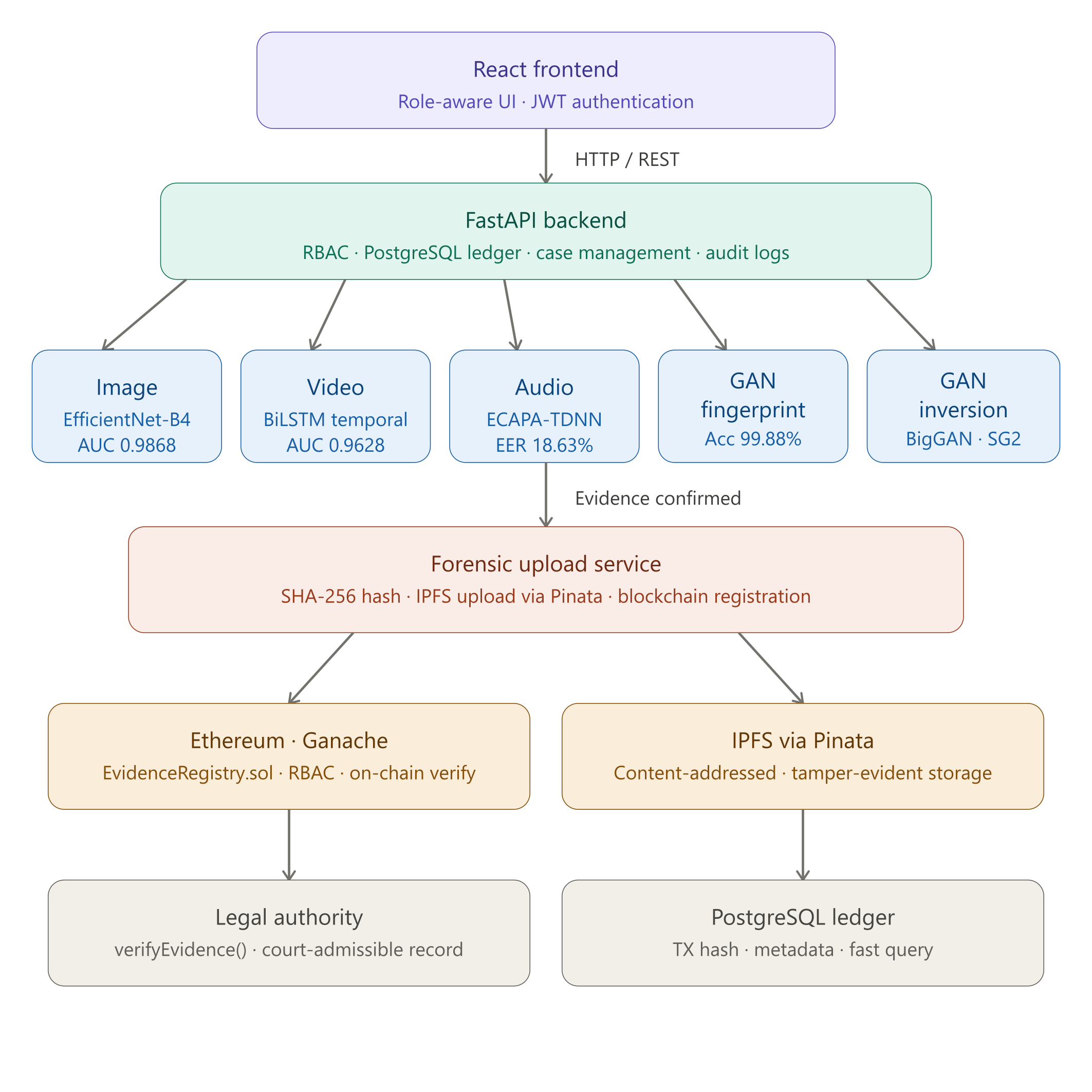}
\caption{High-level system architecture of DeepFake Forensics AI. 
Detection requests flow from the React frontend through the FastAPI 
backend to modality-specific AI modules. Confirmed evidence is 
anchored on Ethereum via IPFS.}
\label{fig:architecture}
\end{figure}

DeepFake Forensics AI is structured as a microservices platform 
comprising four loosely coupled layers: a React frontend, a FastAPI 
backend, a suite of AI inference modules, and a blockchain evidence 
layer. Figure~\ref{fig:architecture} illustrates the high-level 
system architecture.

\subsection{Frontend}

The React frontend provides a role-aware interface 
supporting four user roles: \texttt{FORENSIC\_ANALYST}, 
\texttt{LEGAL\_AUTHORITY}, \texttt{ADMIN}, and 
\texttt{NORMAL\_USER}. Forensic analysts submit media 
files for detection and initiate blockchain evidence 
registration. Legal authorities verify registered 
evidence on-chain. Administrators manage user accounts 
and audit logs. Normal users access a simplified 
dashboard for binary real/fake detection across all 
supported modalities, without access to blockchain 
or case management features.

\subsection{Backend}

The FastAPI backend exposes a RESTful API with JWT authentication 
and role-based access control (RBAC). Separate routers handle 
image detection, video detection, audio detection, GAN 
fingerprinting, GAN reconstruction, forensic upload, and court 
verification. SQLAlchemy manages a PostgreSQL ledger that records 
all detection events, evidence hashes, and case metadata.

\subsection{AI Inference Layer}

The AI layer comprises four independent inference modules, each 
serving a distinct detection task:

\begin{itemize}
    \item \textbf{Image detector}: EfficientNet-B4 binary classifier 
    operating on individual frames.
    \item \textbf{Video detector}: EfficientNet-B4 feature extractor 
    followed by a Bidirectional LSTM temporal aggregator.
    \item \textbf{Audio detector}: ECAPA-TDNN operating on 
    80-dimensional log-Mel spectrogram features.
    \item \textbf{GAN fingerprinter}: Two-stage pipeline comprising 
    a binary real/fake classifier followed by a four-class GAN 
    architecture identifier, both using a custom residual CNN with 
    Squeeze-and-Excitation blocks and a high-pass filter preprocessing 
    layer.
\end{itemize}

A GAN inversion module additionally reconstructs the latent 
representation of fake images using BigGAN~\cite{brock2018biggan} 
and StyleGAN2~\cite{karras2020stylegan2} latent projectors, 
providing evidence of generative origin.

\subsection{Blockchain Evidence Layer}

When a forensic analyst uploads evidence, the system computes a 
SHA-256 hash of the raw file bytes and uploads the file to IPFS 
via the Pinata SDK, receiving a content identifier (CID). The 
hash and CID are then submitted to the \texttt{EvidenceRegistry} 
Solidity smart contract deployed on a local Ethereum network 
(Ganache). The contract stores the hash, CID, evidence type, 
analyst address, and block timestamp. A \texttt{verifyEvidence} =
function allows legal authorities to mark evidence as 
court-verified, emitting an on-chain event for auditability. 
Duplicate submissions are rejected at the contract level via a 
mapping guard, preventing hash collisions. The transaction hash 
and metadata are additionally persisted in the PostgreSQL ledger 
for fast querying without requiring on-chain reads.

\section{Methodology}
\label{sec:method}

\subsection{Image Deepfake Detection}

Frame-level deepfake detection is formulated as a binary 
classification task. We fine-tune an EfficientNet-B4 
backbone~\cite{tan2019efficientnet} pretrained on ImageNet, 
replacing the classification head with a two-layer MLP 
(512 hidden units, GELU activation, 0.3 dropout) followed 
by a 2-class softmax output. Input frames are resized to 
$380 \times 380$ pixels and normalised with ImageNet 
statistics. Training uses AdamW with cosine annealing 
($lr = 10^{-4}$, $T_{max} = 30$) and mixed-precision 
training on FaceForensics++ c23 frames split into 
train/val/test partitions. Class-weighted cross-entropy 
addresses the natural imbalance between real and fake 
frames in the dataset.

\subsection{Video Deepfake Detection}

Video-level detection operates in two stages. First, 
the EfficientNet-B4 frame model extracts 1,792-dimensional 
feature embeddings from uniformly sampled frames. These 
embeddings are stored as per-video \texttt{.npy} files. 
Second, a Bidirectional LSTM temporal model ingests 
sequences of 32 frame embeddings and produces a binary 
real/fake prediction. The BiLSTM uses 512 hidden units 
per direction across 2 layers, followed by a classification 
head with LayerNorm, GELU activation, and 0.4 dropout. 
Training uses \texttt{BCEWithLogitsLoss} with AdamW 
and \texttt{ReduceLROnPlateau} scheduling. The model 
is evaluated on Celeb-DF v2~\cite{li2020celeb} embeddings 
extracted using the trained frame model.

\subsection{Audio Deepfake Detection}

Audio detection uses ECAPA-TDNN~\cite{desplanques2020ecapa}, 
a temporal dilated neural network with emphasis on channel 
attention and multi-scale feature aggregation. Input 
waveforms are converted to 80-dimensional log-Mel 
spectrograms (window 400, hop 160, $f_{min}$=20Hz, 
$f_{max}$=7600Hz) and normalised per utterance. The 
ECAPA-TDNN encoder produces 192-dimensional speaker 
embeddings via attentive statistics pooling. A linear 
classifier with AAMSoftmax loss is used during training 
on ASVspoof2019 LA~\cite{liu2019asvspoof}, with the 
linear projection replaced by a standard softmax 
classifier at inference.

\subsection{GAN Fingerprinting}

The GAN fingerprinting pipeline operates in two stages, 
using images drawn from the GenImage benchmark~\cite{zhu2023genimage} 
covering four generative architectures: ADM, BigGAN, Glide, and VQDM.
Stage 1 classifies images as real or GAN-generated using 
a binary classifier. Stage 2 identifies the specific 
generative architecture among four classes: ADM, BigGAN, 
Glide, and VQDM. Both stages share the same backbone: 
a custom residual CNN with Squeeze-and-Excitation 
(SE) blocks~\cite{hu2018squeeze} and a high-pass filter 
(HPF) preprocessing layer. The HPF, implemented as a 
fixed $3 \times 3$ Laplacian convolution, amplifies 
spectral artefacts characteristic of different GAN 
architectures. The backbone processes grayscale inputs 
through three residual stages with SE attention, 
followed by global average pooling and a linear 
classifier. Both models are trained with AdamW and 
cosine annealing with early stopping on validation 
accuracy.

\subsection{GAN Inversion}

For fake images identified by the fingerprinter, the 
system optionally reconstructs the latent vector 
responsible for their generation. BigGAN~\cite{brock2018biggan} 
inversion projects an input image into the BigGAN 
latent space via iterative optimisation. StyleGAN2~\cite{karras2020stylegan2} 
inversion uses the official latent projector. The 
reconstructed latent vector and its nearest neighbours 
in the generator's learned manifold constitute forensic 
evidence of generative origin, usable in legal proceedings 
to establish the sophistication and likely tooling 
of the forgery.

\subsection{Blockchain Evidence Registry}

Evidence registration follows a deterministic four-step 
protocol. First, the raw file bytes are hashed with 
SHA-256 to produce a content fingerprint. Second, 
the file is uploaded to IPFS via the Pinata SDK, 
returning a content identifier (CID) that 
cryptographically binds the file to its location. 
Third, \texttt{EvidenceRegistry.registerEvidence(hash, CID, type)} 
is called on the deployed Solidity contract via 
Web3.py, storing the hash, CID, evidence type, 
analyst Ethereum address, and block timestamp 
on-chain. Fourth, the transaction hash and metadata 
are persisted in PostgreSQL for fast querying. 
The contract enforces duplicate rejection via a 
mapping guard and emits auditable events on both 
registration and verification. Legal authorities 
invoke \texttt{verifyEvidence(hash)} to mark 
evidence as court-verified, creating an immutable 
on-chain record of the verification event.

\section{Experiments and Results}
\label{sec:results}

\subsection{Experimental Setup}

All models were trained from scratch on a single NVIDIA 
GPU. Table~\ref{tab:datasets} summarises the datasets 
used for training and evaluation.

\begin{table}[htbp]
\caption{Datasets Used for Training and Evaluation}
\label{tab:datasets}
\centering
\begin{tabular}{llr}
\toprule
\textbf{Model} & \textbf{Dataset} & \textbf{Test Samples} \\
\midrule
Image detector   & FaceForensics++ (c23) & 21,211 frames \\
Video detector   & Celeb-DF v2           & 6,529 videos  \\
Audio detector   & ASVspoof2019 LA       & 71,237 utterances \\
GAN fingerprinter & GenImage   & 4,800 images  \\
\bottomrule
\end{tabular}
\end{table}

\subsection{Image Deepfake Detection}

The EfficientNet-B4 image detector achieved 92.57\% 
accuracy and AUC = 0.9868 on 21,211 held-out frames 
from FaceForensics++. The model demonstrated high 
precision on fake detection (0.9952), with only 81 
false negatives on real samples, reflecting a 
conservative detection bias appropriate for forensic 
applications. The class imbalance in the test set 
(3,068 real vs. 18,143 fake frames) reflects the 
natural composition of the FaceForensics++ benchmark.

\begin{figure}[htbp]
\centering
\begin{subfigure}{0.48\columnwidth}
\includegraphics[width=\textwidth]{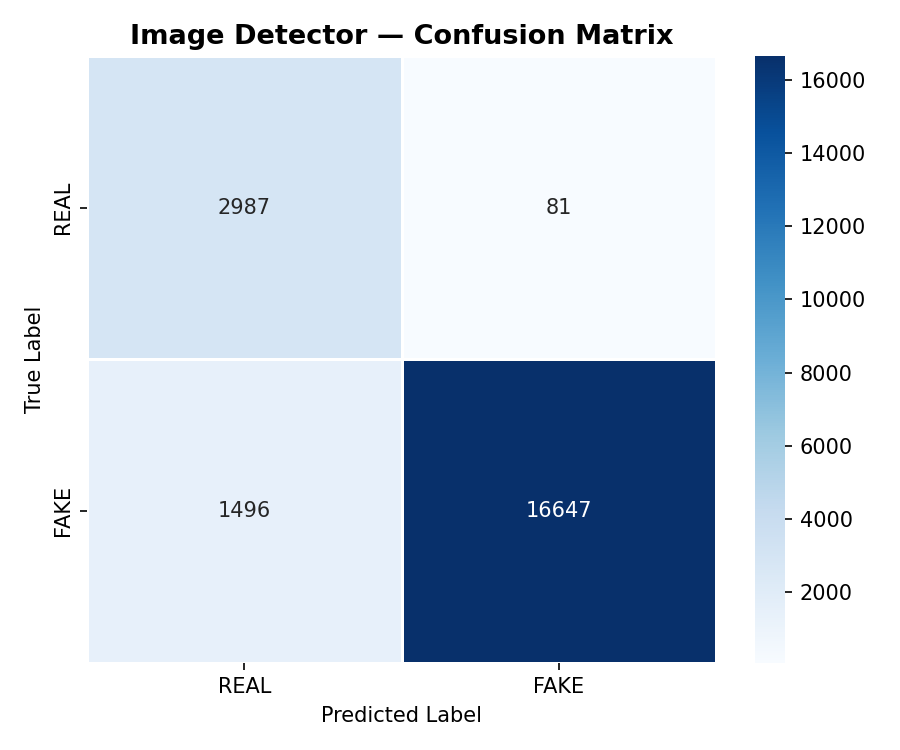}
\caption{Confusion matrix}
\end{subfigure}
\hfill
\begin{subfigure}{0.48\columnwidth}
\includegraphics[width=\textwidth]{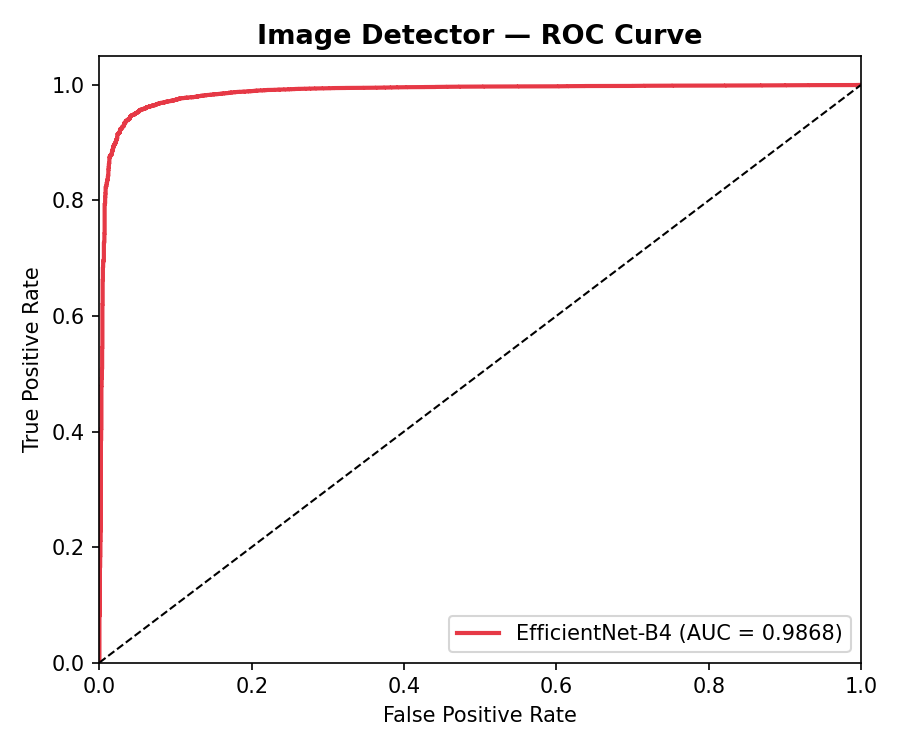}
\caption{ROC curve}
\end{subfigure}
\caption{Image deepfake detector evaluation on FaceForensics++ test set.}
\label{fig:image_results}
\end{figure}

\subsection{Video Deepfake Detection}

The Bidirectional LSTM temporal model achieved 95.54\% 
accuracy and AUC = 0.9628 on 6,529 videos from 
Celeb-DF v2. Operating on 32-frame EfficientNet-B4 
embedding sequences, the model correctly classified 
5,560 of 5,639 fake videos and 678 of 890 real 
videos, yielding a macro-averaged F1-score of 0.8989. 
The lower recall on real videos (0.7618) is consistent 
with the severe class imbalance in the dataset 
(1:6.3 real-to-fake ratio) and represents an 
appropriate operating point for forensic use, 
where false negatives on fakes are the more 
critical error.

\begin{figure}[htbp]
\centering
\begin{subfigure}{0.48\columnwidth}
\includegraphics[width=\textwidth]{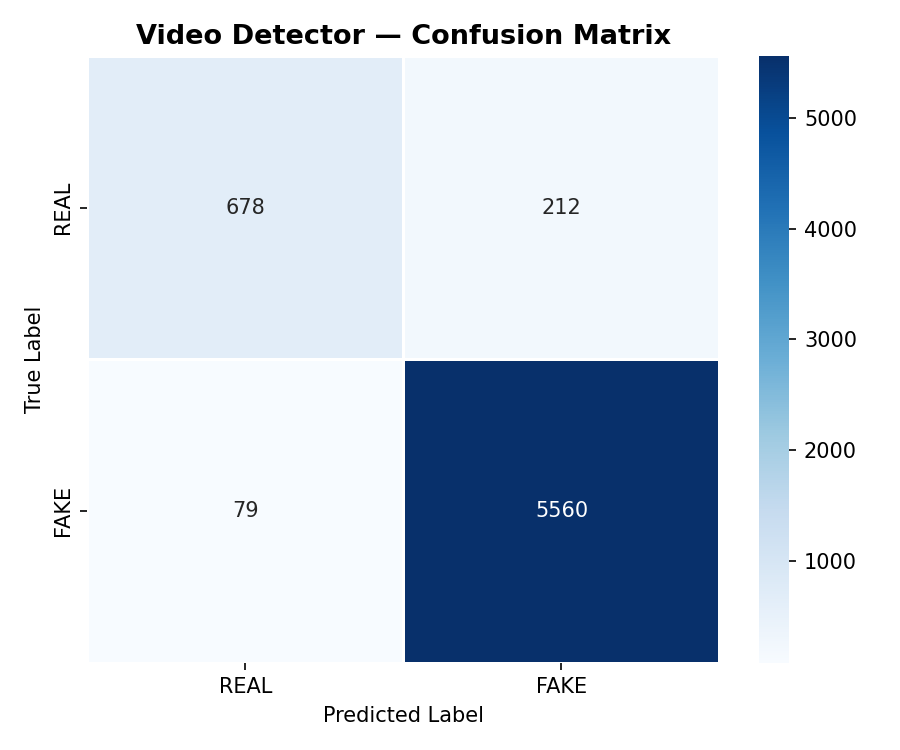}
\caption{Confusion matrix}
\end{subfigure}
\hfill
\begin{subfigure}{0.48\columnwidth}
\includegraphics[width=\textwidth]{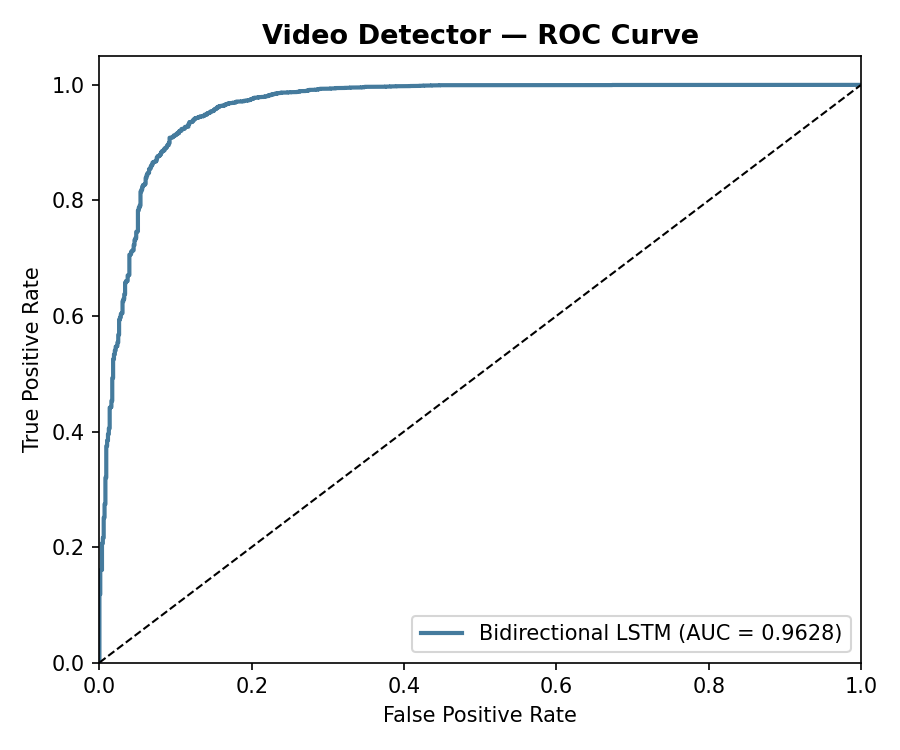}
\caption{ROC curve}
\end{subfigure}
\caption{Video temporal detector evaluation on Celeb-DF v2.}
\label{fig:video_results}
\end{figure}

\subsection{Audio Deepfake Detection}

The ECAPA-TDNN audio detector achieved 87.99\% 
accuracy, AUC = 0.8508, EER = 18.63\%, and 
minDCF = 0.0100 on the ASVspoof2019 LA evaluation 
set of 71,237 utterances. The low minDCF indicates 
strong spoof detection capability under realistic 
operating conditions. The EER reflects the challenge 
of the task under limited fine-tuning epochs and 
the severe class imbalance of the ASVspoof2019 LA 
evaluation partition (1:8.7 bonafide-to-spoof ratio).

\begin{figure}[htbp]
\centering
\begin{subfigure}{0.48\columnwidth}
\includegraphics[width=\textwidth]{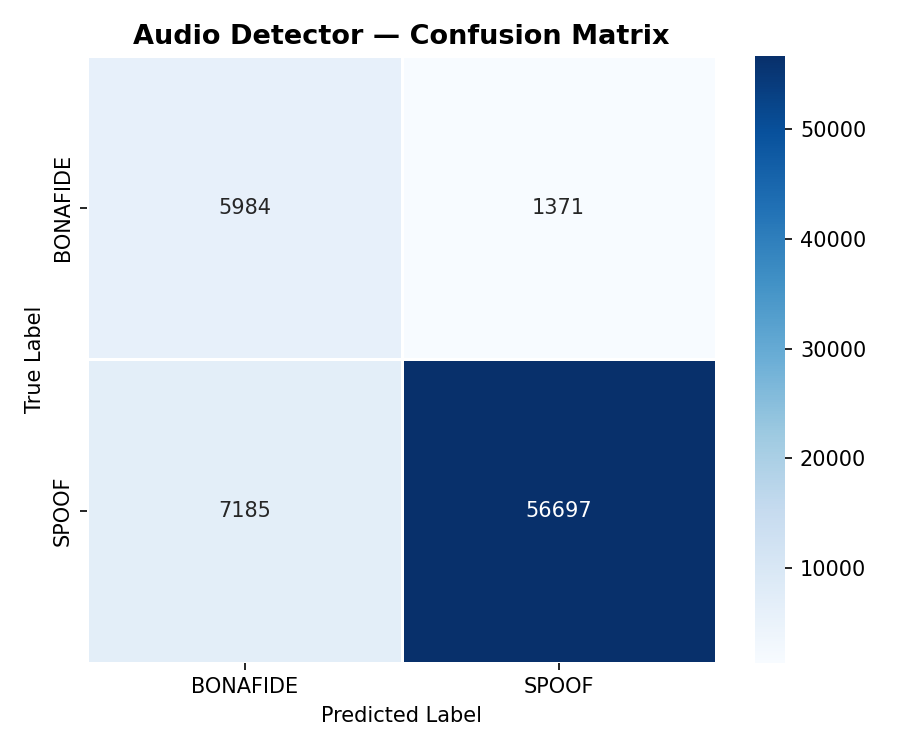}
\caption{Confusion matrix}
\end{subfigure}
\hfill
\begin{subfigure}{0.48\columnwidth}
\includegraphics[width=\textwidth]{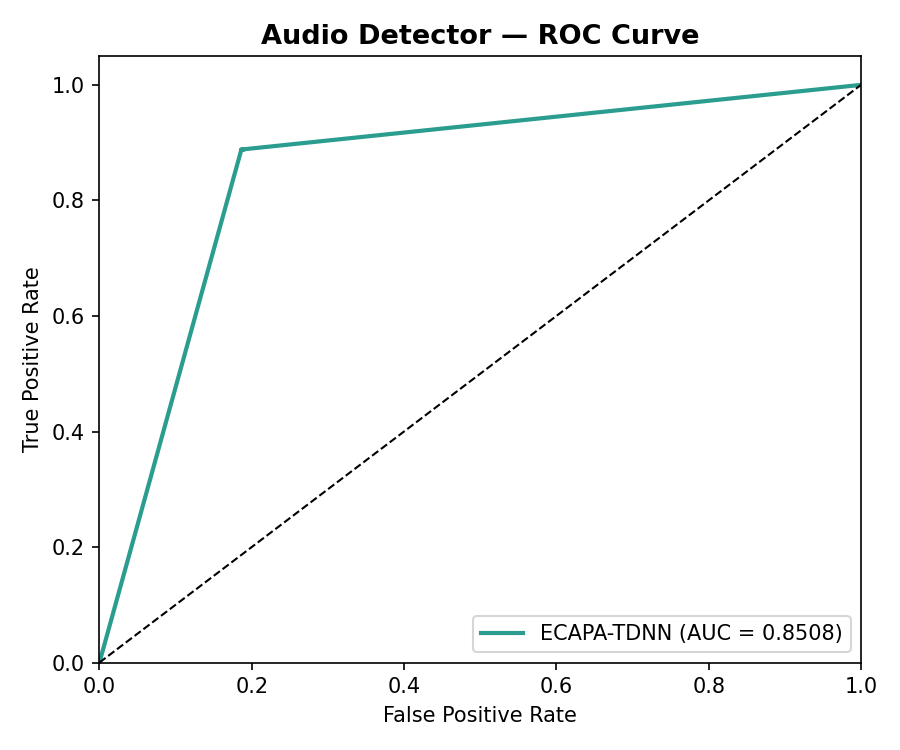}
\caption{ROC curve}
\end{subfigure}
\caption{Audio deepfake detector (ECAPA-TDNN) on ASVspoof2019 LA eval set.}
\label{fig:audio_results}
\end{figure}

\subsection{GAN Fingerprinting}

The GAN fingerprinting module achieved 99.88\% 
overall accuracy on a held-out test set of 4,800 
images spanning four generative architectures. 
Table~\ref{tab:gan_results} reports per-class 
precision, recall, and F1-score. BigGAN was 
classified perfectly (F1 = 1.000). Only 6 images 
were misclassified across the entire test set, 
all between architecturally similar diffusion 
models (ADM and Glide). All four classes achieved 
AUC = 1.000 under one-vs-rest evaluation.

\begin{table}[htbp]
\caption{GAN Fingerprinter Per-Class Results (4,800 test images)}
\label{tab:gan_results}
\centering
\begin{tabular}{lrrr}
\toprule
\textbf{Class} & \textbf{Precision} & \textbf{Recall} & \textbf{F1} \\
\midrule
ADM    & 1.0000 & 0.9992 & 0.9996 \\
BigGAN & 1.0000 & 1.0000 & 1.0000 \\
Glide  & 1.0000 & 0.9958 & 0.9979 \\
VQDM   & 0.9950 & 1.0000 & 0.9975 \\
\midrule
\textbf{Macro avg} & 0.9988 & 0.9988 & 0.9988 \\
\bottomrule
\end{tabular}
\end{table}

\begin{figure}[htbp]
\centering
\begin{subfigure}{0.48\columnwidth}
\includegraphics[width=\textwidth]{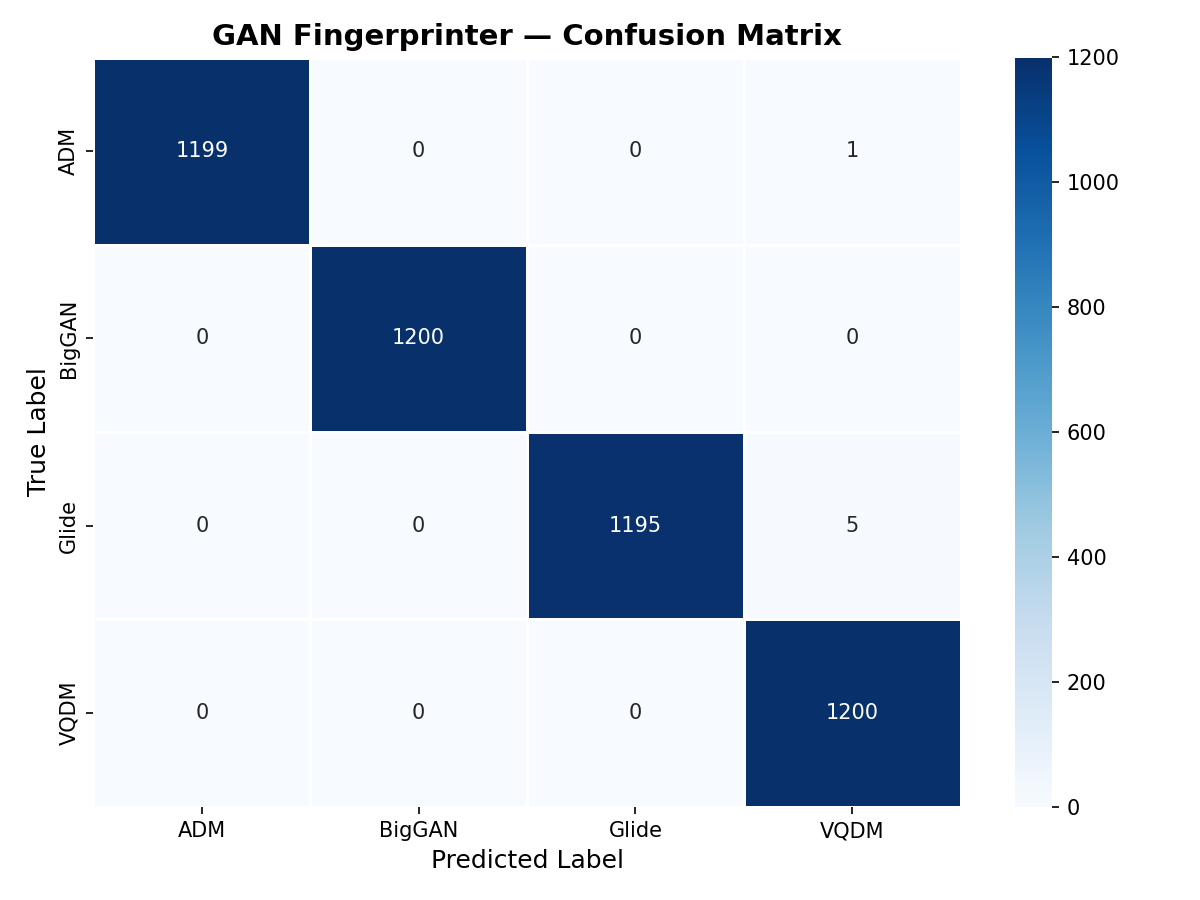}
\caption{Confusion matrix}
\end{subfigure}
\hfill
\begin{subfigure}{0.48\columnwidth}
\includegraphics[width=\textwidth]{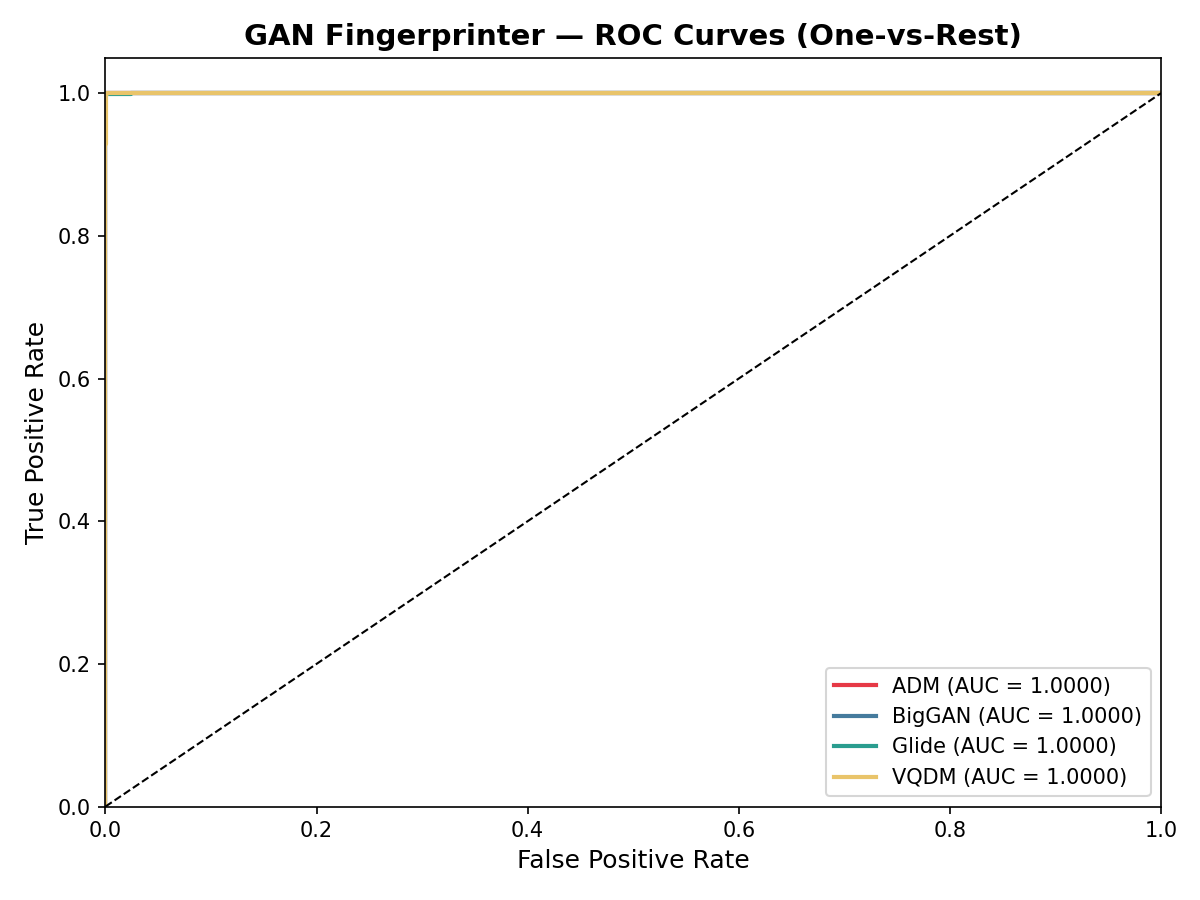}
\caption{ROC curves (one-vs-rest)}
\end{subfigure}
\caption{GAN fingerprinter evaluation across four generative architectures.}
\label{fig:gan_results}
\end{figure}

\subsection{Summary}

Table~\ref{tab:summary} summarises the performance 
of all four detection modules.

\begin{table}[htbp]
\caption{Summary of Detection Module Performance}
\label{tab:summary}
\centering
\begin{tabular}{llrr}
\toprule
\textbf{Module} & \textbf{Backbone} & \textbf{Accuracy} & \textbf{AUC} \\
\midrule
Image detector    & EfficientNet-B4  & 92.57\% & 0.9868 \\
Video detector    & BiLSTM           & 95.54\% & 0.9628 \\
Audio detector    & ECAPA-TDNN       & 87.99\% & 0.8508 \\
GAN fingerprinter & Residual CNN+SE  & 99.88\% & 1.0000 \\
\bottomrule
\end{tabular}
\end{table}

\section{Conclusion}
\label{sec:conclusion}

We presented DeepFake Forensics AI, a unified platform 
that addresses two critical gaps in the current deepfake 
detection landscape: the absence of multi-modal detection 
coverage and the lack of tamper-proof evidence preservation 
for legal proceedings.

Our system trains four independent neural networks from 
scratch, achieving strong performance across all modalities: 
92.57\% accuracy (AUC = 0.9868) for frame-level image 
detection on FaceForensics++, 95.54\% accuracy 
(AUC = 0.9628) for video-level temporal detection on 
Celeb-DF v2, 87.99\% accuracy (EER = 18.63\%) for 
audio spoof detection on ASVspoof2019 LA, and 99.88\% 
accuracy (AUC = 1.000) for GAN architecture fingerprinting 
across four generative models. To our knowledge, the 
combination of multi-modal detection with GAN fingerprinting, 
latent space reconstruction, and blockchain-anchored 
chain-of-custody management represents a novel contribution 
to the field of AI forensics.

Several directions remain for future work. First, the 
audio detector would benefit from class-balanced training 
and additional epochs to reduce the EER below 10\%. 
Second, the blockchain layer currently runs on a local 
Ganache testnet; production deployment would require 
migration to a public Ethereum testnet (Sepolia) or a 
permissioned chain such as Hyperledger Fabric. Third, 
real-time inference optimisation via model quantisation 
or knowledge distillation would make the system viable 
for live video stream analysis. Fourth, cross-dataset 
generalisation --- training on FaceForensics++ and 
evaluating on unseen manipulation types --- remains 
an open challenge that future iterations of this 
platform will address.

As synthetic media generation continues to advance, 
the forensic infrastructure to authenticate digital 
evidence must advance alongside it. DeepFake Forensics 
AI represents a step toward a complete, legally 
admissible forensic workflow for AI-generated media 
detection.

\bibliographystyle{IEEEtran}
\bibliography{references}

\end{document}